\documentclass{article}

\usepackage{PRIMEarxiv}

\usepackage[utf8]{inputenc} 
\usepackage[T1]{fontenc}    
\usepackage{hyperref}       
\usepackage{url}            
\usepackage{booktabs}       
\usepackage{amsmath}        
\usepackage{amsfonts}       
\usepackage{nicefrac}       
\usepackage{microtype}      
\usepackage{fancyhdr}       
\usepackage{graphicx}       
\usepackage{caption}        
\usepackage{etoolbox}       
\usepackage{tikz}
\graphicspath{{figures/}}

\definecolor{gsblue}{HTML}{4285F4}
\DeclareRobustCommand{\gsicon}{%
	\begin{tikzpicture}[baseline=-0.35em]
	\draw[gsblue, fill=gsblue] (0,0) circle [radius=0.16];
	\node[white, font=\bfseries\scriptsize] at (0,0) {G};
	\end{tikzpicture}%
}

\newcommand{\scholarauthorA}{vU6oBhwAAAAJ}  
\newcommand{\scholarauthorB}{ZF8aLrUAAAAJ}  
\newcommand{\scholarA}{\hspace{0.25em}\href{https://scholar.google.com/citations?user=\scholarauthorA}{\gsicon}}
\newcommand{\scholarB}{\hspace{0.25em}\href{https://scholar.google.com/citations?user=\scholarauthorB}{\gsicon}}

\hypersetup{
    colorlinks=true,
    linkcolor=black,
    urlcolor=blue,
    citecolor=black,
    pdfborder={0 0 0}
}

\title{FusionCut: Boundary Representation (B-Rep) Based and Cloud-Ready Cutter Workpiece Engagement (CWE) for Virtual Machining
\thanks{\textit{\underline{Citation}}:
\textbf{H.S. Bank, N.B. Bugdayci. FusionCut: B-Rep Based and Cloud-Ready CWE for Virtual Machining. Proceedings of the 2026 IISE Annual Conference.}}
}

\author{
  H. Sinan Bank\scholarA\thanks{Corresponding author: \texttt{sinan.bank@colostate.edu}} \\
  Department of Systems Engineering \\
  Colorado State University \\
  Fort Collins, USA \\
  \texttt{sinan.bank@colostate.edu} \\
  \And
  N. Bircan Bugdayci\scholarB \\
  Department of Mechanical Engineering \\
  Michigan State University \\
  East Lansing, USA \\
  \texttt{bugdayci@msu.edu} \\
}

\begin{document}
\maketitle

\begin{abstract}
Cutter-workpiece engagement (CWE) is the instantaneous contact geometry between the cutter and the in-process workpiece, playing a fundamental role in machining process simulation and directly affecting the prediction of cutting forces and process stability. The difficulty and challenge of CWE determination come from the complexity of continuously changing geometry, especially for multi-axis milling. To fulfill the requirement of generality---for any cutter type, workpiece shape, or toolpath---the research community has largely pursued two paths: geometrically exact solid modeling and approximate discrete modeling. The former, while accurate, has been hampered by reliance on proprietary, inaccessible software, hindering reproducibility and collaborative research. The latter sacrifices geometric fidelity for algorithmic generality, often leading to computational trade-offs.

This paper presents a framework, FusionCut, that leverages the Boundary Representation (B-Rep) solid modeling kernel of an accessible, modern CAD/CAM platform Autodesk Fusion 360---as freely available for educational and non-commercial use. Our objective is to provide a reproducible framework for the B-Rep approach, while challenging the prevailing assumption that discrete methods such as the triangle meshes are required for general-purpose applications. By providing an accessible implementation and testing it with publicly available models and experiments, we aim to establish a baseline for what is computationally feasible and scientifically necessary for high-fidelity virtual machining. FusionCut offers a path to democratize advanced machining simulation, fostering a more open and progressive scientific ecosystem in digital manufacturing.
\end{abstract}


\section{Introduction}
High-fidelity simulation is a cornerstone of modern digital manufacturing, enabling the creation of ``digital twins'' that can predict and optimize physical processes before chips are ever cut \cite{lee2023digital}. At the heart of virtual machining lies the geometric modeling of the process, where the accurate determination of Cutter-Workpiece Engagement (CWE)---the instantaneous contact geometry between the tool and workpiece---is paramount. The quality of the CWE calculation directly dictates the fidelity of downstream physical models that predict cutting forces, chatter stability, and thermal effects. An error or inefficiency at this foundational stage compromises the entire predictive chain.

\begin{figure}[h!]
\centering
\includegraphics[width=\textwidth]{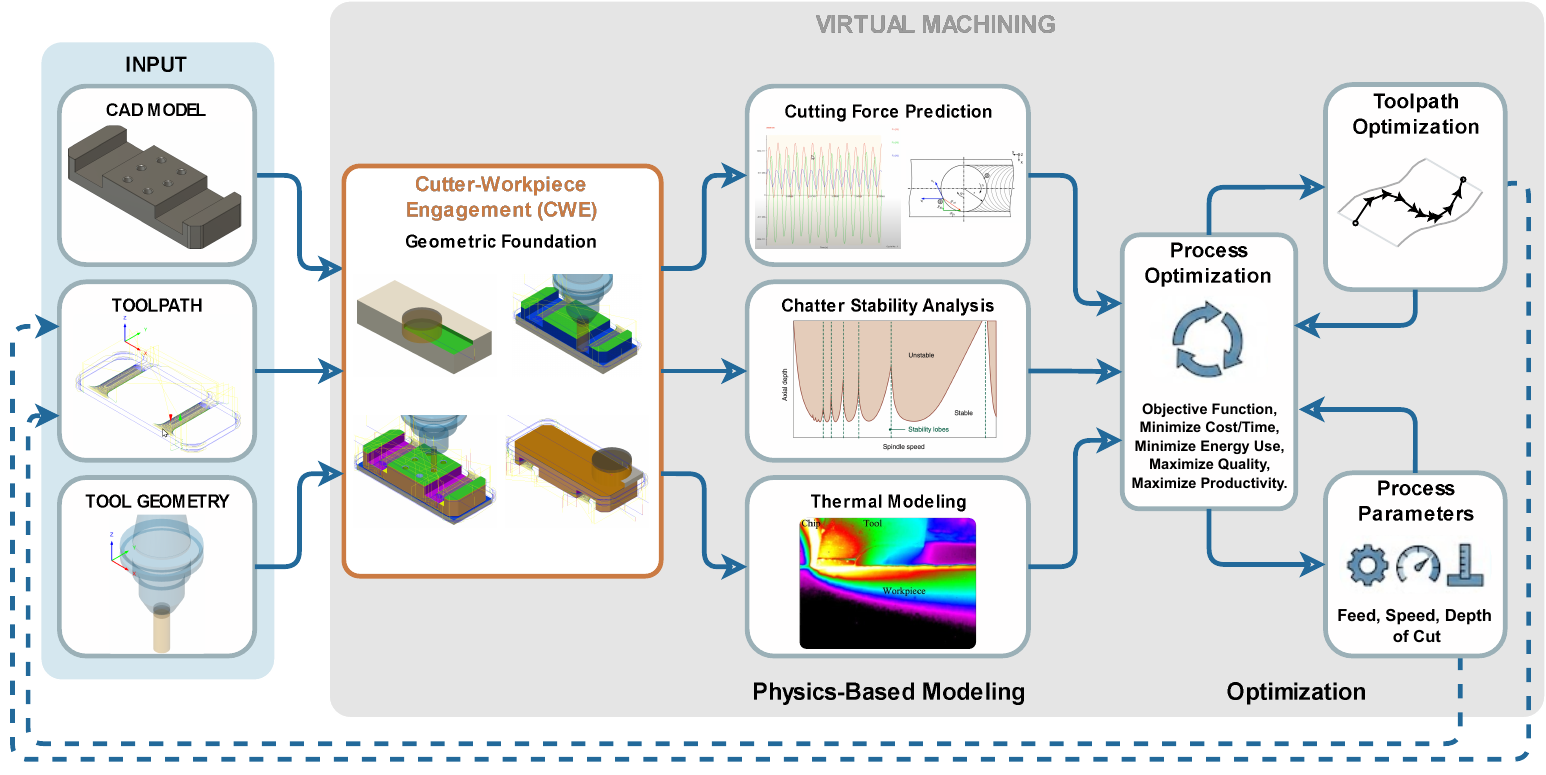}
\caption{Virtual machining workflow and the Cutting-Workpiece Engagement (CWE)'s role.}
\label{fig:workflow}
\end{figure}

The central challenge in CWE modeling is fundamentally a computational representation problem of the geometry with continuously changing aspects of the cutter and workpiece, especially for multi-axis milling. To fulfill the requirement of generality---for any cutter type, workpiece shape, or toolpath---the research community has largely pursued two paths: geometrically exact solid modeling and approximate discrete modeling. However, the field has been constrained by a fundamental trade-off between geometric accuracy and accessibility, creating a critical gap that hinders reproducible research and collaborative advancement in virtual machining.

This paper introduces FusionCut to resolve this long-standing impasse. By utilizing the API of Autodesk Fusion 360, we aim to democratize access to B-Rep-based CWE simulation and challenge the necessity of approximation. This approach demonstrates that commercial-grade solid modeling kernels can be effectively harnessed for research, establishing a fully reproducible and transparent benchmark \cite{fusioncutCodeData2025}. To ensure complete end-to-end reproducibility, all validation is performed using publicly available, standardized models and manufacturing inputs \cite{buildingblocks2025}. By combining an accessible tool with open data, we provide a robust foundation for the future of virtual machining research, fostering a more open scientific ecosystem and providing a foundational tool for model-based systems engineering in digital manufacturing.

\section{Related Work}

The determination of cutter-workpiece engagement in multi-axis milling has been addressed through two fundamentally different methodological approaches, each with distinct advantages and limitations. The first approach, based on Boundary Representation (B-Rep) solid modeling, represents geometry as mathematically exact solids and employs Boolean operations to simulate material removal by subtracting the tool's swept volume from the workpiece \cite{yigit2015solid}. This method, often implemented using commercial geometric kernels such as Parasolid, offers unparalleled geometric fidelity and is widely regarded as the ``ground truth'' benchmark against which all other CWE methods are measured \cite{boz2015comparison}. The solid modeler approach demonstrates better accuracy for complex geometries, particularly in applications requiring precise engagement angle calculations for cutting force prediction and chatter stability analysis \cite{lazoglu2011five,altintas2012manufacturing}. However, the most powerful implementations of this methodology have historically been developed within proprietary, closed-source systems, creating significant barriers to reproducibility, validation, and collaborative research advancement.

In response to the accessibility limitations of solid modeling approaches, the research community has developed a diverse family of discrete, approximate methods. These include voxel-based representations, Z-buffer methods, dexel and tri-dexel field approaches, and triangle mesh-based techniques \cite{aras2008cutter}. These discrete methods offer several practical advantages: they can be implemented using open-source geometric libraries without requiring proprietary geometry kernels, are amenable to object-oriented software architecture, and highly suitable for modern hardware acceleration through GPU parallelization \cite{inui2019cutter}. However, these methods inherently sacrifice geometric exactness for computational efficiency. The discrete nature of these representations introduces approximation errors, including staircase effects on curved surfaces, resolution-dependent fluctuations in engagement angles, and potential misinterpretation of contact regions due to discretization artifacts \cite{boz2015comparison}. These geometric inaccuracies can propagate into downstream physics models, compromising the accuracy of cutting force predictions and process stability analyses, particularly in applications requiring high precision such as thin-wall machining or aerospace component manufacturing.

The field has thus been confronted with a fundamental trade-off: pursue geometric perfection through solid modeling methods that remain largely inaccessible to the broader research community, or accept geometric approximation through discrete methods that prioritize computational efficiency and algorithmic generality. This dichotomy highlights a gap in the literature: the absence of a common, accessible platform for high-fidelity CWE simulation that combines the geometric accuracy of B-Rep methods with the reproducibility and transparency necessary for collaborative scientific advancement. This gap runs counter to the long-standing call for a unified ``modeling discipline'' \cite{bodner2002structured} and integrated systems built upon an authoritative geometric foundation \cite{liu2023deforming}. The presented work addresses this gap by introducing FusionCut, a framework that leverages the accessible API of a modern CAD/CAM platform to provide B-Rep-based CWE calculation with full reproducibility and transparency.

\begin{figure}[b!]
    \centering
    \includegraphics[width=\textwidth]{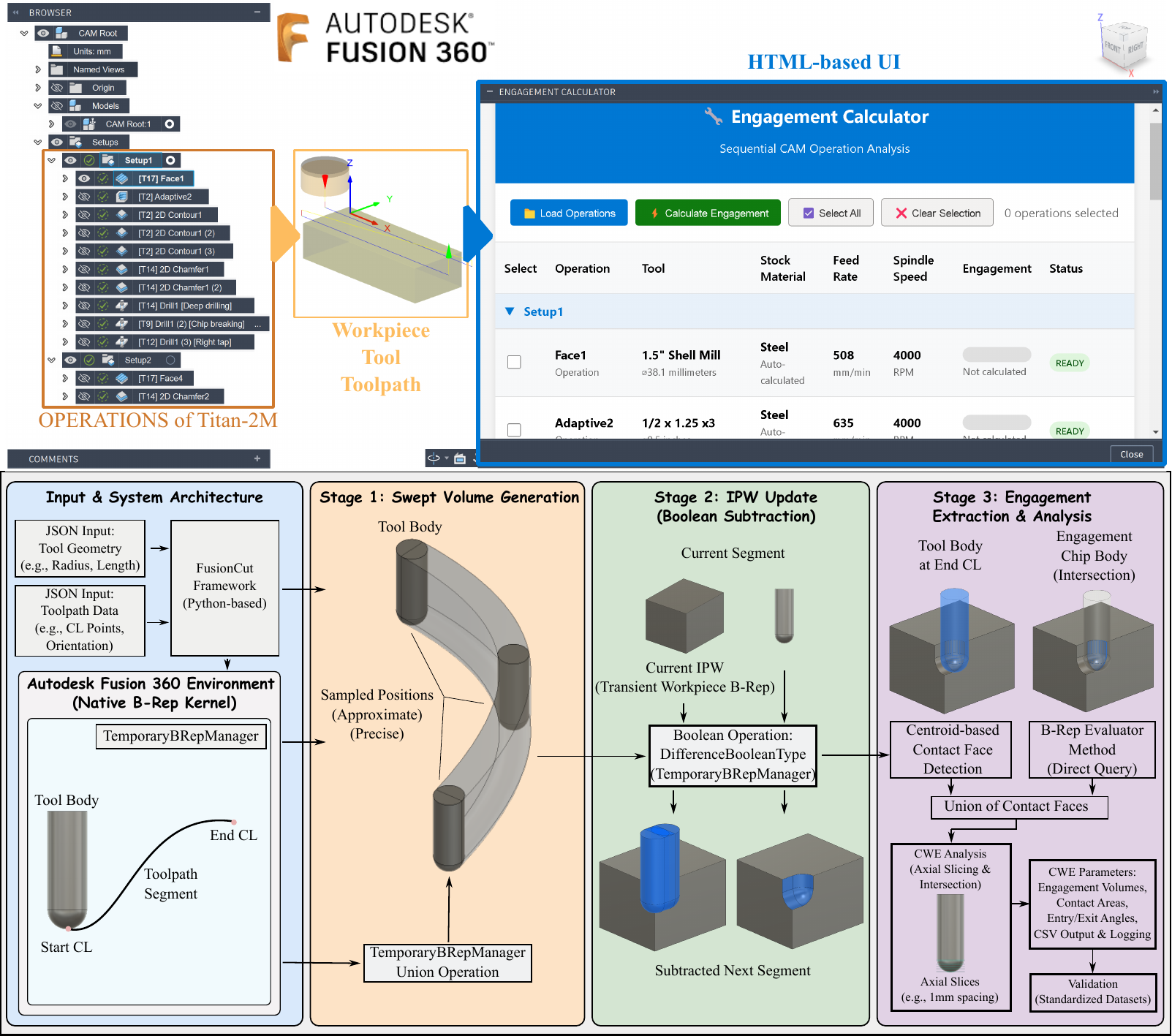}
    \caption{FusionCut cloud-ready frontend and backend framework architecture.}
    \label{fig:architecture}
\end{figure}

\section{Methodology}

FusionCut is implemented as a Python-based framework that leverages the Autodesk Fusion 360 Application Programming Interface (API) to perform B-Rep-based CWE calculation. The framework operates entirely within the Fusion 360 environment, utilizing the platform's native geometric modeling kernel for all solid modeling operations. The system accepts tool geometry definitions and toolpath data in JSON format, processes cutter locations sequentially along the toolpath, and outputs CWE parameters including engagement volumes, contact areas, and engagement angles for each cutter location point. The key architectural advantage of this approach is that it operates directly on the same B-Rep geometric kernel used by the CAD/CAM platform, ensuring geometric consistency and eliminating the need for external geometric libraries.

The core CWE calculation algorithm follows a three-stage process for each toolpath segment between consecutive cutter locations. In the first stage, swept volume generation, the framework creates a transient B-Rep solid in memory representing the material removal envelope as the tool moves from one cutter location to the next. The implementation employs a precise sampling strategy that positions the tool body at multiple intermediate positions along the path segment, with sampling density determined by the tool radius to ensure complete coverage without gaps. All sampled tool positions are then combined using th1e Fusion 360 TemporaryBRepManager to create a single continuous swept volume solid that accurately represents the tool's motion envelope.

The second stage, in-process workpiece (IPW) update, employs Boolean subtraction operations to remove the swept volume from the current workpiece state. The framework maintains a transient B-Rep body representing the evolving workpiece geometry throughout the simulation. At each toolpath segment, the swept volume is subtracted from this transient workpiece using the TemporaryBRepManager's booleanOperation method with DifferenceBooleanType, creating an updated IPW that reflects the material removal from the current segment. This updated workpiece state is then used as the input for the subsequent segment, enabling sequential processing of the entire toolpath while maintaining geometric accuracy through exact B-Rep operations.

The third stage, engagement extraction and analysis, determines the instantaneous CWE at the end cutter location of each segment. The framework first computes the intersection between the tool body positioned at the end cutter location and the updated IPW, creating an engagement chip body that represents the material in contact with the tool. Contact face detection is performed using two complementary methods: a centroid-based approach that identifies faces whose centroids lie on the tool surface, and a B-Rep evaluator method that directly queries face geometry for contact with the tool body. The union of faces identified by both methods ensures comprehensive contact detection. The CWE analysis then extracts engagement parameters by performing axial slicing of the contact geometry. The framework creates a series of planes perpendicular to the tool axis, spaced at configurable intervals (e.g., 1 mm and 0.2 mm), and intersects these planes with the contact faces to determine entry and exit angles at each axial height. This slicing approach produces a complete engagement profile including total contact area, engagement volume, and angular engagement boundaries that can be directly used for cutting force prediction models.

To ensure reproducibility and transparency, all validation test cases use publicly available manufacturing datasets from the Titans of CNC Academy \cite{buildingblocks2025}, which provides open-access CAD models and toolpath data. All input data files, including tool definitions, toolpaths, and stock geometries, are provided in JSON format from the cloud-ready frontend (as shown in Figure~\ref{fig:architecture}), enabling complete reproducibility of the simulation results. The framework outputs comprehensive CSV files containing CWE parameters for each cutter location, along with detailed logging of all geometric operations, allowing for full traceability of the calculation process. All experiments were conducted using Autodesk Fusion 360 (version 2605.1.52) on a system with an NVIDIA RTX 4090 Mobile GPU, with memory consumption ranging from 2--4 GB and GPU utilization between 5--15\% during processing. The complete implementation and validation datasets \cite{fusioncutCodeData2025} are publicly available to ensure full reproducibility.

\begin{table}[htbp]
    \centering
    \caption{FusionCut computational performance for Titan-2M \cite{buildingblocks2025} (top: 1~mm and bottom: 0.2~mm axial sampling).}
    \label{tab:performance}
    \begin{tabular}{lccc}
    \hline
    \textbf{Part/Operations} & \textbf{\# of Proc. CLs/ \# of CLs} & \textbf{Tot. Proc. Sim. Time (s)} & \textbf{Avg. Time per Proc. CL Seg. (ms)} \\
    \hline
    Titan-2M Setup 1 & 19,460/ 36,695 & 5912.9~s & 303.85~ms \\
    \quad Face1 & 352/ 361 & 59.1~s & 168.37~ms \\
    \quad Adaptive2 & 18,320/ 35,097 & 5,580~s & 304.6~ms \\
    \quad 2D Contour1 & 388/ 451 & 139~s & 308.89~ms \\
    \quad 2D Contour1 (2) & 258/ 535 & 89~s & 346.3~ms \\
    \quad 2D Contour1 (3) & 142/ 251 & 45.8~s & 324.82~ms \\
    \hline
    Titan-2M Setup 1 & 19,460/ 36,695 & 10,073~s & 517.62~ms \\
    \quad Face1 & 352/ 361 & 131~s & 373.21~ms \\
    \quad Adaptive2 & 18,320/ 35,097 & 9,480~s & 517.5~ms \\
    \quad 2D Contour1 & 388/ 451 & 270~s & 600~ms  \\
    \quad 2D Contour1 (2) & 258/ 535 & 123~s & 478.6~ms \\
    \quad 2D Contour1 (3) & 142/ 251 &  69~s&  489.4~ms\\
    \end{tabular}
\end{table}

\section{Results and Discussion}

Table~\ref{tab:performance} presents the computational performance metrics for FusionCut across the test cases at two axial sampling intervals (1~mm and 0.2~mm). The table reports both scheduled cutter locations (\# of CLs) and the number of processed CL segments (\# of Proc. CLs) used in the per-segment timing averages. For Titan-2M Setup 1, FusionCut achieves average times per processed CL segment of 168.37--346.3~ms at 1~mm sampling and 373.21--600~ms at 0.2~mm sampling, with total simulation times of 5912.9~s (1~mm) and 10{,}073~s (0.2~mm). At the 0.2~mm sampling interval, which matches the resolution used in the literature comparison \cite{boz2015comparison}, the 2D contour operations yield 478.6--600~ms per processed segment, which is on the same order as the dexelfield timing reported by Boz et al. \cite{boz2015comparison} (approximately 528~ms per CL segments for 2.5-axis operations---527 segments for 278~s), though direct comparison is limited by differences in geometries, toolpaths, and measurement definitions. Notably, FusionCut successfully handles the complex Adaptive2 toolpath with 35{,}097 scheduled CLs (18{,}320 processed CL segments), demonstrating scalability to industry-grade toolpaths requiring engagement evaluation across large toolpath datasets. While current version of FusionCut's performance is slower than Parasolid-based implementations, this difference is expected given the different implementation architectures (Fusion 360 API versus direct Parasolid kernel access), utilization of different programming languages (Python vs C++), and the fact that our tests benefit from more recent and advanced hardware platforms. Nonetheless, the total simulation time of FusionCut for complex parts remains manageable, demonstrating that high-fidelity B-Rep-based CWE simulation is computationally competitive with approximate discrete methods on accessible platforms.

\begin{figure}[b]
    \centering
    \includegraphics[width=\textwidth]{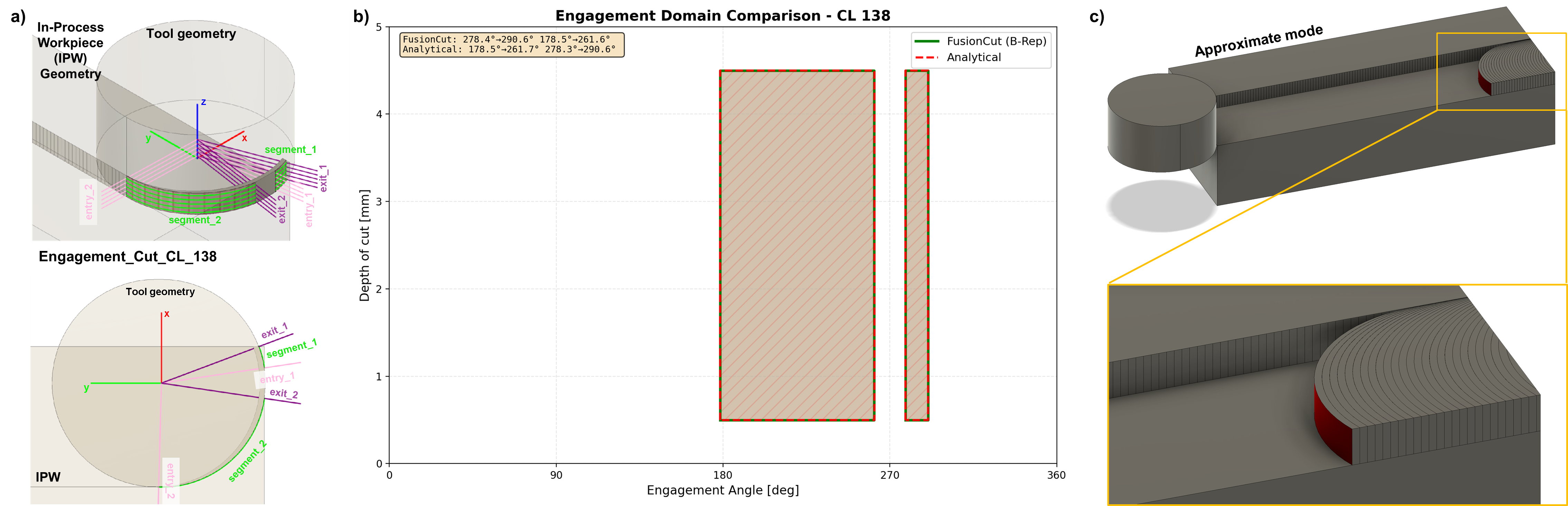}
    \caption{(a) 3D/2D CWE, (b) Engagement domain for analytical validation, (c) Virtually machined part.}
    \label{fig:fidelity}
\end{figure}

The geometric fidelity of FusionCut is demonstrated through validation on complex machining operations. Fig.~\ref{fig:fidelity} presents a multi-panel visualization of CWE geometry validation via the first operation (face milling) of Titan-2M Setup 1. Top part of Fig.~\ref{fig:fidelity}(a) shows a 3D isometric view of the tool geometry, in-process workpiece (IPW), and engagement segments with entry/exit lines, illustrating the instantaneous contact geometry between the cutter and workpiece. Bottom section of Fig.~\ref{fig:fidelity}(a) provides a 2D top-down view showing the engagement angles and engagement segments on the tool circumference, demonstrating the precise angular characterization of the CWE. Fig.~\ref{fig:fidelity}(b) displays a 3D view of the virtually machined part with zoomed sections highlighting geometric details. The engagement segments and angles shown in Fig.~\ref{fig:fidelity}(a) demonstrate the exact geometric representation achieved through B-Rep, which contrasts with the approximation artifacts that characterize discrete methods, particularly in applications involving complex free-form surfaces \cite{frank2002cnc}. This visualization illustrates the capability of B-Rep-based methods to provide geometrically exact CWE characterization for virtual machining applications.

To further validate the geometric accuracy of FusionCut's engagement calculations, we compared the computed engagement angles against an independent analytical solution based on trigonometric circle-line and circle-circle intersection calculations. Fig.~\ref{fig:fidelity}(b) presents this comparison for CL 138, a corner case where the tool exits the stock boundary, resulting in two distinct engagement areas. The analytical model accounts for the in-process workpiece geometry by modeling the stock as a rectangle with material removed by the stadium-shaped tool swept volume from previous cutter locations. As shown in the figure, FusionCut's B-Rep-based results (green) closely match the analytical solution (red hatched), with angular differences within 0.25\textdegree. For simple mid-pass cases, the agreement is within 0.01--0.02\textdegree, effectively validating the trigonometric correctness of FusionCut's engagement angle extraction. This analytical validation provides independent confirmation that the B-Rep-based approach produces geometrically accurate CWE characterization.

Synthesizing the findings from computational performance and geometric fidelity results, FusionCut successfully achieves the geometric fidelity of traditional B-Rep methods with practical computational performance, all on an accessible and reproducible platform. This work provides a viable solution to the fidelity-versus-accessibility trade-off that has long defined the field. The combination of exact geometric representation with practical computational performance demonstrates that the dichotomy between accuracy and accessibility is not fundamental, but rather a consequence of implementation choices. By leveraging modern, accessible CAD/CAM technology, we have shown that high-fidelity virtual machining simulation can be both geometrically accurate and widely accessible, enabling a new generation of reproducible research in digital manufacturing.

\section{Conclusion and Future Work}

This paper introduced FusionCut, a reproducible framework for B-Rep-based CWE simulation that addresses a critical gap in virtual machining research: the lack of accessible, high-fidelity geometric simulation tools. The primary scientific contribution of this work is demonstrating that modern CAD/CAM platform APIs can serve as viable foundations for research-grade CWE computation, providing an alternative path between inaccessible proprietary kernels and geometrically approximate discrete methods. By combining an accessible platform (Autodesk Fusion 360) with publicly available datasets (Titans of CNC Academy) and open-source implementation, we establish a reproducible baseline that enables the research community to verify, extend, and build upon our results---addressing the reproducibility challenges in this field.

The current work has several limitations that define directions for future research. First, while FusionCut demonstrates competitive performance with discrete dexelfield methods, direct comparison with literature results is constrained by differences in test geometries, toolpaths, and hardware configurations. Future work will establish standardized benchmark cases with complete geometric and toolpath specifications to enable rigorous cross-method comparisons. Second, the framework currently processes cutter locations sequentially; cloud-based parallelization will be explored to improve scalability for large industrial toolpaths. Third, a C++ implementation will be developed to reduce API overhead and enable more direct performance comparisons with optimized implementations. Fourth, experimental validation through cutting force measurements will be conducted to verify that the geometric accuracy of FusionCut translates to improved physical predictions. Finally, the methodology will be extended to support 5-axis toolpaths and complex tool geometries, enabling validation across a broader range of industrial applications. These efforts aim to advance virtual machining research toward the broader goal of creating true-to-life digital twins that can accurately predict and optimize manufacturing processes \cite{lee2023digital}.

\section*{Acknowledgments}
This is an initial draft of the paper accepted for presentation at the 2026 IISE Annual Conference \& Expo. The authors gratefully acknowledge the Titans of CNC Academy (\url{https://academy.titansofcnc.com}) for providing publicly available CAD models and manufacturing data used in the validation of this work.

\bibliographystyle{unsrt}
\bibliography{references}

\end{document}